\documentstyle[12pt]{article}
\catcode`\@=11

\@addtoreset{equation}{section}
\def\theequation{\arabic{section}.\arabic{equation}}

\catcode`\@=11

\def\section{\@startsection{section}{1}{\z@}{3.5ex plus 1ex minus
   .2ex}{2.3ex plus .2ex}{\large\bf}}

\def\eqnarray{\let\@currentlabel=\theequation\refstepcounter{equation}
    \global\@eqnswtrue
    \global\@eqcnt\z@\tabskip\@centering\let\\=\@eqncr
  $$\halign to \displaywidth\bgroup\@eqnsel\hskip\@centering
   $\displaystyle\tabskip\z@{##}$&\global\@eqcnt\@ne
      \hfil${{}##{}}$\hfil
      &\global\@eqcnt\tw@ $\displaystyle\tabskip\z@{##}$\hfil
       \tabskip\@centering&\llap{##}\tabskip\z@\cr}

 \def\lefteqn#1{\hbox to 4\arraycolsep{$\displaystyle #1$\hss}}

\def\harr#1{\smash{\mathop{\hbox to .5in{\rightarrowfill}}
  \limits^{ #1}}}
\def\varr#1{\llap{$\scriptstyle #1$}\left\downarrow}
\def\varr2#1{\llap{$\scriptstyle #1$}\left\uparrow}
\def\partz#1{\displaystyle {\partial {#1} \over {\partial z}}}
\def\partzbar#1{\displaystyle {\partial {#1} \over {\partial {\bar z}}}}

\newcommand{\req}[1]{(\ref{#1})}
\newcommand{\be}{\begin{equation}}
\newcommand{\ee}{\end{equation}}
\newcommand{\bea}{\begin{eqnarray}}
\newcommand{\eea}{\end{eqnarray}}
\def\Det{\rm Det}

\def\IR{{\hbox{{\rm I}\kern-.2em\hbox{\rm R}}}}
\def\IH{{\hbox{{\rm I}\kern-.2em\hbox{\rm H}}}}
\def\IC{{\ \hbox{{\rm I}\kern-.6em\hbox{\bf C}}}}
\def\IZ{{\hbox{{\rm Z}\kern-.4em\hbox{\rm Z}}}}

\begin{document}
\begin{titlepage}
\vspace{.5in}
\begin{flushright}
UCD-95-37\\
August 1995\\
hep-th/9510192 \\
\end{flushright}
\vspace{.5in}
\begin{center}
{\large\bf Quantum Liouville Theory from a Diffeomorphism \\[1ex]
 Chern-Simons Action}\\

\vspace{.4in}
M.~ C.\ A{\sc shworth} \footnote{\it email: mikea@landau.ucdavis.edu}\\
{\small\it Department of Physics}\\
{\small\it University of California}\\
{\small\it Davis, California 95616 USA}\\
\end{center}

\vspace{.5in}
\begin{center}
{\bf Abstract}
\end{center}
\begin{center}
\begin{minipage}{5in}
{\small
A Chern-Simons action written with Christoffel Symbols has a natural
gauge symmetry
of diffeomorphisms.  This Chern-Simons action will induce a
Wess-Zumio-Witten model on the boundary of the manifold.  If we restrict
the diffeomorphisms to chiral diffeomorphism, the Wess-Zumio-Witten model
is equivalent to a quantum Liouville action.
}
\end{minipage}
\end{center}
\end{titlepage}
\addtocounter{footnote}{-1}

\section{Introduction}

Over the years, much work has gone into studying Liouville theory.
In addition to being a natural candidate for two dimensional gravity,
Liouville theory also arises naturally in non-critical dimensions in
string theory. Some attempts have been made to further
understand this theory by studying Chern-Simons theories on manifolds with
boundary.
Some of this work has been done in studies of topologically massive
gravity \cite{carl} \cite{kogan}.  Topologically massive gravity contains
a Chern-Simons term that is written in terms of spin connections.
These spin connections
possess an SO(2,1) gauge symmetry.
This Chern-Simons action will
induce an SO(2,1) Wess-Zumino-Witten (WZW)
model on the boundary of the manifold.
SO(2,1) is then homomorphic to SL(2,\IR),
and SL(2,\IR) WZW is known to contain
Liouville theory \cite{oguri} \cite{alex} \cite{forg}.
Thus, from a three dimensional gravity theory, we get a two
dimensional gravity theory induced on the boundary of the manifold.

In this paper, I would like to extend this method in the following way.
It has been shown that the Chern-Simons action expressed in terms of
spin connections is equivalent to a Chern Simons action
expressed in terms ofChristoffel symbols \cite{perc}
(For simplicity I shall call this diffeomorphism Chern-Simons theory).
Using this information as a starting point,
I would like to complete the following diagram.

 $$
 \matrix{
 SO(2,1)\ \hbox{\em CS} & \harr{}  & \hbox{\em SL} (2,\IR)\ \hbox{\em WZW}
 & \harr{} & \hbox{\em Liouville} \cr
 \downarrow  &&&&  \uparrow {\scriptstyle ?} \cr
 \hbox{\em Diffeo\  CS}  && \harr{?} && \hbox{\em Diffeo\ WZW} \cr
  }
 $$

\section{WZW from Chern-Simons Theory}

I will start with the following Chern-Simons action on a 3-dimension
manifold M with metric h.

\be
{\bf S}_{CS} = {k \over {4 \pi}} \int_{\rm M}
{ d^3 x \ \varepsilon ^ {\mu \nu \lambda}
\left( \Gamma _ {\mu \ b} ^{\ a}  \partial _ {\nu} \Gamma _ {\lambda \ a}
^ {\ b} - {2 \over 3}  \Gamma _ {\mu \ b} ^{\ a} \Gamma _ {\nu \ c} ^ {\ b}
 \Gamma _ {\lambda \ a} ^{\ c} \right)}
\ee

\noindent
Following Percacci's paper \cite{perc},  $ \Gamma _ {\mu \ b} ^{\ a} $ is
defined in the following way.
Let the metric tensor be written as

\be
h_{\mu \nu} = \theta ^a _{\ \mu} \theta ^b _{\ \nu}
\kappa _{a b}.
\label{eq1}
\ee

\noindent
Then from (\ref{eq1}), we see that

\bea
\Gamma _{\lambda \ b} ^{\ a} & = & \theta ^e _{\ \lambda}
\kappa ^{a f} \Gamma _{e f b}
\nonumber \\
\Gamma _{e f b} & = & { 1 \over 2} \left(
\theta ^{ \ \mu} _ b \partial _{\mu} \kappa _{ e f}
+ \theta ^{ \ \mu} _ e \partial _{\mu} \kappa _{ f b}
- \theta ^{ \ \mu} _ f \partial _{\mu} \kappa _{ b e} \right)
+ {1 \over 2} \left( c _{e f b} + c _{f b e} - c _{b e f} \right)
\nonumber \\
 c_{efb} & = & { \theta _{e \lambda}} \left( {\theta _f ^{\ \mu}
\partial _{\mu}  \theta _b ^{\ \lambda} - \theta _b ^{\ \mu}
\partial _{\mu}  \theta _f ^{\ \lambda}} \right) .
\label{eq2}
\eea

\noindent
For a fixed metric, Percacci shows that different values of
$\kappa$ and $\theta$
result in equivalent theories, up to a gauge transformation.
So, if we choose $\kappa _{a b}= \delta _{a b}$ then
from \req{eq1}, we see that $\theta^a _{\ \mu} $
becomes the triad and $\Gamma _{\mu \ b} ^{\ a}$
becomes the spin connection.  If we choose
$ \theta ^a _{\ \mu} = \delta ^a _{\ \mu}$ then from \req{eq1} and \req{eq2},
we see that $ \Gamma _{\mu \ b} ^{\ a}$ are the Christoffel symbols.
We will consider the above action in the latter choice
or ``metric gauge.''  This is the first step in completing the
above diagram.

The natural gauge transformations for this action
are the diffeomorphisms acting on the Christoffel symbols.
The Christoffel symbol transforms as.

\be
\Gamma' (x') _ {a \ b} ^{\ c} = g _{\ f} ^{c}
\  \Gamma (x) _{e\  g} ^{\ f}\  ({g ^{-1}}) _{\ a} ^{e} \  ({g ^{-1}})
_{\ c} ^{g} -  ({g^{-1}}) _{\ a} ^{e}\ \partial _{d} ({g_{\ e} ^{b}})
\ ({g^{-1}})_{\ a} ^{e} ,
\label{gtran}
\ee

\noindent
where $g$ is defined as

\be
(g) _{\ a}^{e} \ = {\partial {x'} ^{e} \over \partial {x} ^{a}}.
\label{gelem}
\ee

\noindent
Now, for simplicity, I will define $\Gamma ^{a}_{\ b} =
\Gamma ^{\ a} _{c \ b}\  dx^{c}$. The initial action now becomes

\be
{\bf S}_{1} = {k \over {4 \pi}} \int_{\rm M} {
\left( \Gamma _ {\ b} ^{a}\ \wedge \  {\rm d}  \Gamma _ {\ a}^ {b}
- {2 \over 3}  \Gamma _ {\ b} ^{a}\ \wedge \ \Gamma _ {\ c} ^ {b}
\ \wedge \ \Gamma _ {\ a} ^{c} \right)} .
\ee

\noindent
The above gauge transformation \req{gtran} now appears as a normal gauge
transformation.

\be
\Gamma' (x') _ {\ b} ^{c} = ({g ^{-1}})_{\ e} ^{b}
\  \Gamma (x) _{\ f} ^{e} \  g _{\ c} ^{f}
 +  ({g^{-1}})_{\ e} ^{b}\  {\rm d}( g) _{\ a} ^{e}
\ee

\noindent
The above action transforms under this gauge transformation as

\be
{\bf S}_1 ( \Gamma ' ) = {\bf S}_1 ({\Gamma}) -
{k \over 4 \pi} \int _{\partial {\rm M}} {\rm Tr} \left( \left(
{\rm d}g g^{-1} \right) \wedge {\Gamma} \right) -
{k \over 12 \pi} \int _{\rm M} {\rm Tr} \left( g ^{-1} {\rm d}g
\right) ^3
\ee

If the manifold has no boundary the second term is zero, and the last term
is just the winding number.  Thus up to the winding number, this action on
a manifold without boundary is invariant under this guage transformation.
However, with the boundary, if we take the variational derivative
to find the classical equation of motion, a boundary term also appears.

\be
\delta {\bf S}_1 = {k \over 2 \pi} \int _{\rm M} {\rm Tr}
\left( \delta \Gamma
\left( {\rm d} \Gamma + \Gamma \wedge \Gamma \right) \right) -
{k \over 4 \pi} \int _{\partial {\rm M}} {\rm Tr} \left( \Gamma
\wedge \delta \Gamma \right)
\ee

\noindent
To be able to find the extremum for the classical equation of motion,
the boundary terms must be zero.  We would like it
to be zero without fixing all of the fields on the boundary or setting them
equal to zero. This problem is equivalent to picking the necessary boundary
data to solve the equations of motion.  A standard choice in this system is
to pick a complex structure on the boundary, and fix one component
of the field, and then add a surface term to the original action
to cancel the above boundary term.
This choice of field fixing allows us to uniquely map solutions of
the classical equations of motion on
one manifold to another across a shared boundary.
This allows sewing (or gluing) manifolds with boundaries together to create
a new manifold with a well defined action (see \cite{carl2}).
In our case we will choose to fix $\Gamma _{z\  b} ^{\ a} $, and
add the following boundary term for the action.

\be
{\bf S} _2 = {k \over 4 \pi} \int _{\partial {\rm M}} \Gamma _{{\bar z}\  b}
^{\ a}  \Gamma _{z\  a} ^{\ b}  dz d{\bar z}
\ee

Taking the variation of the resulting action we get

\be
\delta ( {{\bf S} _1 + {\bf S} _2} ) =
{k \over 2 \pi} \int _{\rm M} {\rm Tr} \left( \delta \Gamma
\left( {\rm d} \Gamma + \Gamma \wedge \Gamma \right) \right)
 - {k \over 2 \pi} \int _{\partial {\rm M}} \Gamma _{{\bar z}\  b} ^{\ a}
\delta  \Gamma _{z\  a} ^{\ b}  dz d{\bar z}.
\ee

\noindent
$\delta \Gamma _{z\  b} ^{\ a}$
is equal to zero, and the boundary term is zero.
With this ``fixed'' action, the extremum now gives well defined
equations of motion.

Considering the gauge transformation on this new action with the additional
boundary term, we get

\be
({\bf S}_1 + {\bf S}_2) [\Gamma' ] = ({\bf S}_1 + {\bf S}_2) [\Gamma]
+ k {\bf S}^+ _{WZW} [g, \Gamma]
\ee

\be
{\bf S}^+ _{WZW} [g, \Gamma] = {1 \over 4 \pi} \int _{\partial
{\rm M}}  {\rm Tr} \left( g^{-1} \partial _z g g^{-1} \partial _
{\bar z} - 2 g ^{-1} \partial _z g \ \Gamma _{\bar z} \right) +
{1 \over 12 \pi} \int _{\rm M} {\rm Tr} \left( g^{-1} {\rm d} g
\right) ^3.
\ee

I would like to consider this action in a path integral setting.
To begin with, we can use
the standard Fadeev-Popov gauge fixing methods to get
the measure,

\be
[{\rm d} \Gamma'] = [{\rm d} \Gamma] [{\rm d} g]\ \delta\left(
F[\Gamma] \right) \Delta _F [\Gamma] .
\label{chmeas}
\ee
\noindent
$F$ is our gauge fixing function $F(g)=0$.
In the Fadeev-Popov method, $\Delta _F$ must be gauge invariant.
In order for $\Delta _F$ to be gauge invariant $[{\rm d} g]$ must
be a left invariant measure (see Ryder \cite{ryder} for more details).
The path integral now becomes

\be
{\bf Z} = \left( \int [{\rm d} {\Gamma}] \delta \left( F[
{\Gamma} ] \right)  \Delta [{\Gamma}] \exp \left \{
i{\bf S}[{\Gamma}] \right \} \right)
\left( \int [{\rm d} g] \exp \left \{ ik {\bf S}_{WZW}^+
[g,{\Gamma}_{z}]\right \}  \right)
\ee

\noindent
The gauge dependent piece factors out, and because ${\Gamma}_{\bar z}$
is fixed,
these two pieces are independent of each other.  The first term is the
``bulk'' term and can be ignored.
The second term is just a WZW model coupled to a fixed back-ground.

A convenient choice of gauge would be to let $h^{ab} \Gamma _{a\ b}
^{\ c} = 0$.  This is a called the harmonic gauge (see ref \cite{hguage}).
The result of this gauge fixing leads to finding the determinant of the
Lichnerowicz operator.

\be
(g^{ab} ( \nabla _a \nabla _b \delta ^c _d) + {\rm R}_d ^c)
 \delta ^2 (z-z')
\ee

\noindent
However this is not needed to continue the analyzes of the WZW model.

\section{Chiral Diffeomorphism and Liouville Theory}

Following Carlip's paper \cite{carl}, I will consider the
chiral diffeomorphism $z \rightarrow w(z,{\bar z})$ and
${\bar z} \rightarrow {\bar z}$ (note: $ g=g^{-1}$ in Carlip's paper).
Let the metric on the original boundary be give by

\be
{\rm d}s^2 = e^{2 \xi} {\rm d}z {\rm d}{\bar z}.
\ee

\noindent
We see that  $\Gamma _{z}$, the fixed field on $\partial$M, is

\be
\Gamma _{z} = \left( \matrix{
2 \partial _{z} \xi  & 0     \nonumber \cr
0 & 0 \nonumber \cr}
\right).
\ee

Under this chiral diffeomorphism, the group element \req{gelem}
may be written in the following way.

\be
 g  = {\left( \matrix{
\\[2ex]  {\partz{w(z , {\bar z} )}}
  &   {\partzbar{w(z , {\bar z} )}} \nonumber \cr
\     0      &      1        \nonumber \cr
} \right) }
= {\left( \matrix{
\\[2ex]    {e^\phi}  &  \mu \nonumber \cr
   0      &  1 \nonumber \cr
}\right)}
\label{diffg}
\ee

\noindent
Because the partial derivatives commute, we should also include
constraint equation,

\be
{\partzbar{} \left({ \partz{w} } \right) = \partz{} \left({ \partzbar{w}}
\right)} \ \  \Rightarrow \ \  {{\partzbar{}} \left(  \displaystyle{e ^\phi}
\right) = \partz{\mu}}.
\ee

\noindent
This constraint was not included into the action in ref.\cite{carl}, and
thus the resulting field theory included no cosmological constant term.
\bigskip

Substituting \req{diffg} into ${\bf S}^+_{WZW} [g,{ \Gamma} _{z}]$.

\be
g^{-1} {\rm d} g =  {\left( \matrix{
\\[2ex]  {{\rm d} {\phi}}
  &   {-\mu {\rm d}{\phi}} \nonumber \cr
\     0      &      0        \nonumber \cr
} \right) }
\ee

\be
{\rm first \ term} \ \Longrightarrow
\ {\rm Tr}(g^{-1} \partial _z g g^{-1} \partial _{\bar z} g ) =
\partial _z \phi \partial _{\bar z} \phi
\ee

\be
{\rm second \  term} \ \Longrightarrow
\ {\rm Tr}(g^{-1} \partial _z g \Gamma _z) = 4  \partial _{\bar z} \phi
\partial _z \xi
\label{term3}
\ee

\noindent
Integrating by parts, we get
\be
 4  \partial _{\bar z} \phi \partial _z \xi = 4 \phi \partial _{\bar z}
 \partial _z \xi =  \sqrt{g} R \phi.
\ee

\be
{\rm last \ term} \  \Longrightarrow
\ {\rm Tr} (g^{-1} {\rm d} g) =  \partial _i \phi  \partial _j \phi
\partial _k \phi \epsilon ^{ijk} = 0
\ee

\noindent
So all together we get the action,

\be
{\bf S}[\phi] = \int{( \partial _z {\phi} \partial _{\bar z}
{\phi} + \sqrt{g} R \phi)
dz d{\bar{z}}}.
\label{chir}
\ee

The choice of which boundary field is fixed determines how the gauge
fields couple to the curvature $R$.
If we choose to fix $\Gamma _{\bar z}$ with the above choice for the
chiral gauge field, the gauge field would not couple to
the curvature $R$. Instead (\ref{chir}) would be a free field action.

In the above action (\ref{chir}), there is no dependence on $\mu$.
However, we will see that there is $\mu$ dependence in the measure
when we consider the constraint.
We know the measure because it was induced from the
original Chern-Simons measure \req{chmeas}.
As mentioned before, in the standard Fadeev-Popov method, the gauge group
measure must be left invariant.  It can easily be shown that
$[ dg ]= [e^{- \phi} d\phi] [d\mu]$  is left invariant
(where $[e^{- \phi} d\phi] = \Det[e^{- \phi}] [ d\phi]$ is defined as in
Reference \cite{dhoker}).
We then need to include the constraint into the measure in a left
invariant manner. The following is a left invariant form of the
delta function, which we will include in the the measure.

\be
\delta \left({ \partzbar \phi - e^{- \phi} \partz \mu}\right)
\ee

Writing the delta function as a Fourier transform, we find the full left
invariant measure as

\be
[dg]_{left} = [e^{- \phi} d \phi] [d \mu] [d \lambda] \  \exp
\left \{ {\  - i \lambda
\int_{\partial {\rm M}}{(\partial _{\bar z}
\phi - e ^ {- \phi} \partial _z \mu)}} \right \}.
\ee

Now if we integrate out $\mu$, we will get the delta function

\be
\Det[{e^{\phi}}] \ \delta (\partial _z (\lambda e^ {- \phi})) .
\ee
$\Det[{e^{\phi}}]$  cancels with the like term in the original measure.
Solving for when the delta function is non-zero we see that
\be
\lambda = f(\bar z) e^{\phi},
\ee

\noindent
where $f(\bar z)$ is an arbitrary function of $\bar z$.  Substituting this
into the partition function and integrating by parts gives

\be
{\bf Z} = \int{[d \phi] {\rm Det}[\partial _z] [df(\bar z)]\
 \exp \left \{ {\  \int
_{\partial {\rm M}} {(\partial _{\bar z}
f(\bar z)  e^{\phi})}} \right \}  \  \exp \left \{
{\ {\bf S}^[\phi]} \right \} }.
\ee

\bigskip

Now let $\partial _{\bar z} f({\bar z}) = \mu _0 e^{\phi ' ({\bar z})} $
and $\phi \rightarrow \phi + \phi '({\bar z})$.  Then terms in the action
that depend on $\phi '({\bar z}) $ can be written as complete derivatives
including the term coupling $\phi$ and R. This is easiest to see by looking
back to \req{term3}.
The volume term $ Det[\partial _z] [df({\bar z})]$ can be integrated out.
The final partition function is given by

\bea
{\bf Z} & = & \int{[d \phi] e^{S_L [\phi ,\mu _0]}},
\nonumber \\
{\rm where} \hskip .5in
 {\bf S}_L & = & \int{\partial _{\bar z} \phi \partial _{z} \phi +
\mu _0 e^{\phi} + \sqrt{g} R \phi}
\eea
\noindent
which is the Liouville action. So we now have come full circle in the above
diagram.

\section{Discussion}

It is clear that a chiral diffeomorphism is not the full diffeomorphism
``gauge'' group.  The full group is difficult to deal with because it
can not be simply factored like ordinary gauge groups.
Thus, we can not use the standard Polyakov and Wiegman
factorization \cite{poly}
to simplify the math. We can see this in the following way.  Although the
chain rule lets us multiply two elements to get a new element, the
second element below is not in the original coordinate system.

\be
g(z,\bar z) = g_2(u(z,{\bar z}),\bar u(z,{\bar z})) \ g_1(z,\bar z)
\ee

\be
 \matrix{
  &  (u,{\bar u})  & \cr
   &\hbox{\hskip -1.8cm} {^{g_1} \hskip-.2cm \nearrow}  &
    \hbox{\hskip -1.3cm} {\searrow ^{g_2}} \cr
  (z, {\bar z})  & \harr{g} & (w,{\bar w}) \cr
  }
\ee

\noindent
The above forms a semi-direct product because of the dependence of u on
the $g_1$ transformation.  Thus we can not factor it into a
direct product and use a
Gaussian decomposition of the group that is used in standard WZW models.

However, one can look at additional ``conformal'' diffeomorphisms.  Consider
the transformation

\bea
 & g' (z , {\bar z}) = a ({\bar v})  g (u, {\bar u}) b(z)
 \\[1ex]
 & (z , \bar z) \  \harr{b} \ (u , \bar u) \ \harr{g} \ (v , \bar v)
  \harr{a} \  (w , \bar w).
\nonumber
\eea

\noindent
$ g (z, {\bar z} ) $ is
our chiral diffeomorphism given in \req{diffg}.
The $b(z)$ and $a({\bar v})$ terms
simply re-scale the $z$ and ${\bar z}$ coordinates
(ie, conformal transformations) and therefore can be
dealt with easily because they preserve the complex structure.
For the $a ({\bar v})$, we write out the chain rule and get

\be
 g= {\pmatrix{
\\[2ex] \partz{w} & \partzbar{w} \nonumber \cr
0 & \partzbar{\bar w} \nonumber \cr }}
= {\pmatrix{
\\[2ex] e^ {\phi}  & \mu \nonumber \cr
0 & e^ {\psi} \nonumber \cr }}.
\ee

\noindent
The $\psi$ term comes directly form $a({\bar v})$. This term does not
not couple with the curvature term.  Therefore, it is a free field action.
The constraint $\partial _z ( e^ {\psi}) =0 $ then removes it.
For $b(z)$, we will get $\phi \rightarrow \phi + \phi ' (z)$.
Like before, $\phi ' (z)$ term can be written has a total derivative
 and can be dropped from the action.
So this action is invariant in a similar way to
${\bf S}[a({\bar z})\ g\  b(z)] = {\bf S}[g]$ in \cite{poly} and
\cite{kniz}.

Unlike the earlier methods, this method will lead to an action that
will couple to the curvature as well as the cosmological constant.
Perhaps a full treatment of the diffeomorphism group will give more
information about Liouville theory.

\section{Acknowledgments}

I would like to thank Steve Carlip for all of his support and time.
This work was supported by National Science Foundation grant PHY-93-57203
and Department of Energy grant DE-FG03-91ER40674.


\begin{thebibliography}{99}

\bibitem{carl} S.\ Carlip, Nucl.\ Phys.\ {\bf B362} (1991) 111-124.

\bibitem{kogan} I.\ I.\ Kogan, Nucl.\ Phys.\ {\bf B375} (1992) 362-380.

\bibitem{oguri} M.\ Bershadsky and H.\ Ooguri, Comm.\ Math.\ Phys.\
{\bf 126} (1989) 49.

\bibitem{alex} A.\ Alexeev and S.\ Shatashvili, Nucl.\ Phys.\ {\bf B323}
(1989) 719.

\bibitem{forg} P.\ Forgacs, A.\ Wipf, J.\ Balog, L.\ Feher, and L.\ O'
  Raifeartaigh, Phys.\ Lett.\ {\bf B277} (1989) 214.

\bibitem{perc} R. Percacci, Ann.\ of\ Physics {\bf 177} (1987) 27-37.

\bibitem{carl2} S.\ Carlip, M.\ Clements, S.\ DellaPietra, and V.\
  DellaPeitra, Comm. Math. Phys. {\bf 127} (1990) p 253-271.

\bibitem{ryder} L. H. Rider, ``Quantum Field Theory'', Cambridge University
Press, 1985, p. 252-254.

\bibitem{hguage} R. S. Puzio, Class. Quant. Grav. 11 (1994) 2667-2676.

\bibitem{dhoker} E.\ D'Hoker and P.\ S.\ Kurzepza, Mod. Phys. Lett.
{\bf A5} p 1411-1421.

\bibitem{poly} A.\ M.\ Polyakov and P.\ B.\ Wiegman, Phys. Lett {\bf
141B} (1984) p 223.

\bibitem{kniz} V.\ G.\ Knizhnik and A.\ B.\ Zamolodchikov, Nucl. Phys.
{\bf B247} (1984) p 83-103.

\end{thebibliography}
\end{document}